\documentclass[%
reprint,
superscriptaddress,
amsmath,amssymb,
prl
]{revtex4-1}

\usepackage{graphicx}
\usepackage{dcolumn}
\usepackage{bm}
\usepackage{amssymb}
\usepackage[capitalise]{cleveref} 

\usepackage{graphicx, float}

\begin{document}


\title{Structure-property relationships via recovery rheology in viscoelastic materials}

\author{Johnny Ching-Wei Lee}
\affiliation{%
Department of Chemical and Biomolecular Engineering, University of Illinois at Urbana-Champaign, Illinois, USA 61801
}%

\author{Katie M. Weigandt}
\affiliation{%
Center for Neutron Research, National Institute of Standards and
Technology, Gaithersburg, MD, USA
}%

\author{Elizabeth G. Kelley}
\affiliation{%
Center for Neutron Research, National Institute of Standards and
Technology, Gaithersburg, MD, USA
}%

\author{Simon A. Rogers}%
\email{sarogers@illinois.edu}
\affiliation{%
Department of Chemical and Biomolecular Engineering, University of Illinois at Urbana-Champaign, Illinois, USA 61801
}%



\date{\today}

\begin{abstract}
The recoverable strain is shown to correlate to the temporal evolution of microstructure via time-resolved small-angle neutron scattering (SANS) and dynamic shear rheology.
Investigating two distinct polymeric materials of wormlike micelles and fibrin network, we demonstrate that, in addition to the nonlinear structure-property relationships, the shear and normal stress evolution is dictated by the recoverable strain. A distinct sequence of physical processes under large amplitude oscillatory shear (LAOS) is identified that clearly contains information regarding both the steady-state flow curve and the linear-regime frequency sweep, contrary to most interpretations that LAOS responses are either distinct from, or somehow intermediate between the two cases. This work provides a physically-motivated and straightforward path to further explore the structure-property relationships of viscoelastic materials under dynamic flow conditions.

\end{abstract}

\maketitle



A long-standing challenge in understanding the out-of-equilibrium behavior of soft matter is to link microstructural rearrangements and the macroscopic flow properties. We show, via time-resolved rheo-small-angle neutron scattering (rheo-SANS), that such a link can be made, forming nonlinear rheological structure-property relations, by considering the evolution of the recoverable component of the strain.

The simultaneous collection of macroscopic rheological information and particle- or molecular-level data has been used understand the interplay between material constituents that cover a wide range of length scales. Recent examples include the applications of scattering techniques in polymeric materials \cite{Forster2005,LopezBarron:2012fg,Calabrese2018,Wang2017,Gurnon2014,Rogers_2012softmatter,Leheny2015}, confocal microscopy in colloidal systems \cite{Cheng2011,Ghosh2017,Amann2012,Mutch2013,Cheng2012,Sentjabrskaja2015,Vasisht2018,Besseling2007,Westermeier2016,Mutch2016,Smith2007} and simulation methods \cite{Radhakrishnan:2016kf,Hattemer:2015dq,Hou2010,Leishangthem2017,Xu2018a,Koumakis:2013jg}. Despite these efforts, rheological structure-property relations remain incompletely understood. Oscillatory shearing provides a reproducible way to probe viscoelastic behaviors and has been used to study a wide array of soft materials \cite{Cordier2008,Koos2011,Mangal2015,Kokkinis2015,Zhong2016,Boland2016,Gu2018}. Many processing conditions and practical applications of soft materials can be approximated by large amplitude oscillatory shear (LAOS) because it offers independent control of the length and time scales of structural rearrangements. As such, LAOS has been widely adopted as a model transient flow protocol capable of eliciting nonlinear responses \cite{Leblanc2008,Lee2010,Lettinga:2012dw,LopezBarron:2012fg,Koumakis:2013jg,Munster2013,Hall2016,Radhakrishnan:2016kf,Jaspers:2017hr,Perazzo2017,Tapadia:2006ij,Rogers2018}.

Typical mathematically-based descriptions of experimental responses to LAOS are based on Fourier transformation (FT), which represents the complex sequence of processes exhibited by soft materials in the time domain as a sum of harmonic contributions in the frequency domain \cite{Wilhelm:2002dd}, performing an averaging of sorts. Physical processes that take place sequentially over intervals of time shorter than a period of oscillation are therefore not easily discerned via such analysis methods. The lack of a generic understanding of the higher harmonics \cite{Hyun:2011kd} has limited the widespread adoption of these methods in time-resolved molecular-level studies \cite{MinKim2014,Lettinga:2012dw,Rogers_2012softmatter,LopezBarron:2012fg,Eberle:2012ei,Gurnon2014}.

In this letter, we show that the recoverable strain, an often overlooked rheological metric that was first proposed by Weissenberg \cite{Weissenberg:1947ik} and Reiner \cite{Reiner:1958wp}, provides an ideal basis for understanding the complex evolution of the microstructure and the shear and normal stresses of an industrially-relevant entangled solution of wormlike micelles (WLMs) and a biopolymer network of fibrin (see SI for material preparation). In discussing the experimental reality of recovery tests, Lodge noted that typical experiments are constrained in one dimension. Constrained recovery was incorporated into his transient network theory of polymers, which makes a prediction about the relationship between recoverable strain and the first normal stress difference \cite{Lodge:1958iu}. Several studies \cite{MOONEY:1951vs,Lodge:1958iu,BENBOW:1961uh,Wagner:1978va,Laun:1986ep,PRehbinder1954,Muenstedt2008} have adopted this concept to study polymeric and nanocomposite systems, mostly under steady shear or creep flow. The central variable in these studies \cite{Weissenberg:1947ik,Lodge:1958iu,MOONEY:1951vs,BENBOW:1961uh,Wagner:1978va,Laun:1986ep,Reiner:1958wp} is the amount of deformation recovered after the removal of shear stress. Reiner \cite{Reiner:1958wp} elevated the importance of the recoverable deformation, calling it \textit{the strain}, a definition that has not stuck. Despite significant early efforts, measurements of constrained recovery remain limited, and have not, until now, been applied to the study of structure-property relationships under oscillatory shearing.

\begin{figure*}[t]
	\centering
	\includegraphics[width=.95\textwidth]{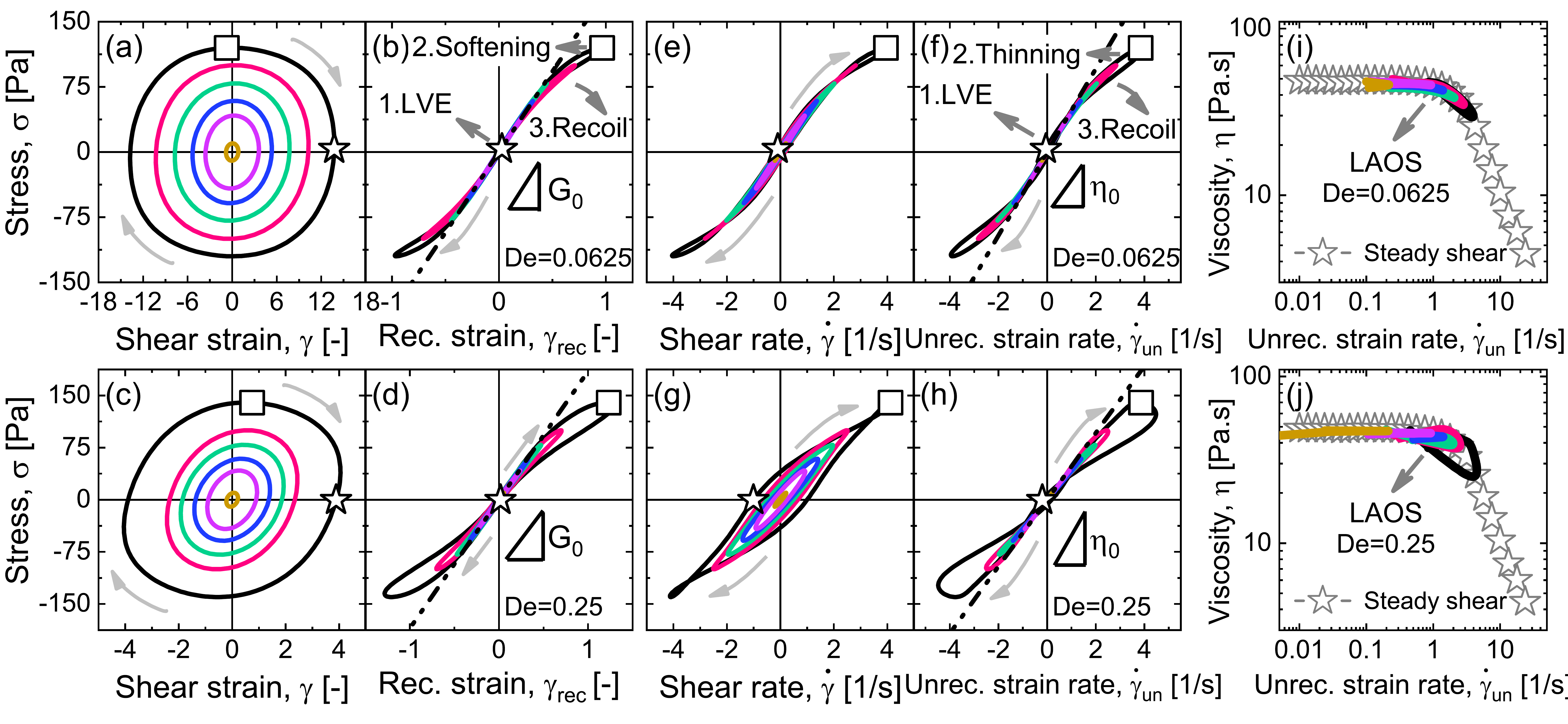}
	\caption{Lissajous curves in the traditional ($\sigma-\gamma$, $\sigma-\dot{\gamma}$) and proposed ($\sigma-\gamma_{rec}$, $\sigma-\dot{\gamma}_{un}$) frames at $De=0.0625$ (a, b, e, f) and $De=0.25$ (c, d, g, h) from WLMs. The lines in the proposed elastic and viscous views have slopes equal to the plateau modulus $G_0=180 Pa$ and the zero-shear viscosity $\eta_0=48$ Pa$\cdot$s. The stars and squares correspond to zero and maximum recoverable strain at the largest amplitude, respectively.
	Shear viscosity $\eta=\sigma/\dot{\gamma}_{un}$ determined from LAOS on top of the steady-shear flow curve at $De=0.0625$ (i) and $De=0.25$ (j).
	}
	\label{Liss}
\end{figure*}

The two polymeric materials we study have very different behaviors, which allows us to more clearly illustrate the benefits of our new approach. Worm-like micelles undergo many breakage and reformation events on the time scale of reptation, leading to a Poisson distribution of length scales at equilibrium, and a single relaxation time \cite{Cates:1990bh}. They have therefore been used to study nonlinear flow behaviors of entangled linear and branched polymers \cite{Spenley:1993bc,Inoue:2005hy,Boukany2008,Forster2005,In1999,Cardiel2013}. Fibrin networks are known to possess flow properties that are distinct from common synthetic polymers \cite{Mackintosh1995,Storm2005,Onck2005,Munster2013} because they are primarily elastic and stiffen when strained, protecting tissue from large deformations. In contrast, synthetic polymer solutions often soften and shear-thin when strained. 

We use in-situ rheo-SANS to simultaneously monitor the alignment of micellar segments as well as the recovery rheology. Measurements are made in the steady alternating state when all initial transience has decayed. An iterative constrained recovery procedure is employed at 200 distinct evenly-spaced instants along an oscillation (see Supplemental Material for detailed experimental protocol, which includes Ref. \cite{Baravian1998}). The unrecovered part of the strain, $\gamma_{un}$, is the strain the system ultimately recovers to, and the recoverable strain, $\gamma_{rec}$, is the part of the total strain that is elastically recovered. 
All rheo-SANS measurements are made with an Anton-Paar MCR-501 rheometer with a concentric-cylinder Couette geometry at the NCNR on beamline NGB-30. Normal stress differences and additional constrained recovery tests are measured using a DHR-3 rheometer (TA Instruments) with a 4-degree cone and plate geometry.

The alignment of micellar segments is measured in the velocity-gradient (1-2) and velocity-vorticity (1-3) planes using time-resolved rheo-SANS techniques \cite{Porcar:2004et,Eberle:2012ei}. A $q$-range of 0.006 1/$\AA$ to 0.03 1/$\AA$ is probed, corresponding to the rod-like scattering of the micelle segments (see SI). SANS data are reduced according to NCNR guidelines \cite{Kline:2006fh}. A temporal deconvolution protocol \cite{Calabrese:2016ie} is used to enhance the resolution and accuracy of the measurement of the underlying structural evolution.

Oscillatory shearing from WLMs with amplitudes ranging from the linear (SAOS) to the nonlinear (LAOS) regimes are performed at Deborah numbers ($De = \omega\lambda$, where $\omega$ is the angular frequency and $\lambda$ is the longest relaxation time) of 0.0625 ($\omega =$0.25rad/s) and 0.25 ($\omega =$1 rad/s), where the inertial effects of the stress-controlled procedure are negligible. 
The traditional elastic and viscous Lissajous curves of the WLMs are displayed in \cref{Liss} (a and e) and (c and g). The distorted elastic Lissajous curve and the secondary loops of the viscous Lissajous curve indicate that the system undergoes a complex sequence of processes over the course of an oscillation.
FT-based analysis schemes quantify the departure from linearity in an average sense, taking the response from an entire period of oscillation as the object to be analyzed. The resulting harmonics therefore have no clear physical relation to the structure at any particular instant. We contrast this with the underlying principle of time-resolved rheo-scattering techniques \cite{Eberle:2012ei,Calabrese:2016ie}, which is that the structure and rheology at any one instant are causally related. A framework that correlates the rheology with real-time structural evolution is needed.

Strain can be decomposed into recoverable and unrecoverable components \cite{Weissenberg:1947ik,Reiner:1958wp}. Recoverable strain is elastic, while viscous properties are dictated by the rate at which strain is acquired unrecoverably. Elastic and viscous Lissajous curves that reflect this view are therefore presentations of stress vs. recoverable strain ($\sigma-\gamma_{rec}$) and stress vs. unrecoverable strain rate ($\sigma-\dot{\gamma}_{un}$). The slopes of the $\sigma-\gamma_{rec}$ and $\sigma-\dot{\gamma}_{un}$ curves for Hookean solids and Newtonian fluids retain their clear meaning of being the modulus and viscosity, respectively. Plots of stress vs. recoverable strain for a Newtonian fluid will have undefined slopes because purely viscous fluids acquire no strain recoverably. Our proposal therefore removes the current possibility of defining a modulus for a purely Newtonian fluid or a viscosity for a purely Hookean solid.

The proposed elastic Lissajous curves, $\sigma-\gamma_{rec}$, from WLMs under LAOS are shown in \cref{Liss}(b and d). The new presentations show significant differences from the traditional plots that use the total strain. At the instants close to the total strain extremes (stars in \cref{Liss} a and c), the recoverable strain is nearly zero, indicating that the material is closest to its equilibrium configuration and is therefore undeformed. Similar behavior has been observed in polymeric and colloidal systems \cite{Rogers:2011hba,Kim:2014dn,Lee:2017gy,Rogers:2017ew,Ramya2019a}, where a constant linear-regime elasticity exists near the strain extrema under LAOS. At both investigated frequencies, the $\sigma-\gamma_{rec}$ curves show straight lines of equal slope at $\mid\gamma_{rec}\mid<\sim$0.5, which indicates an elastic modulus of 180 Pa. This value corresponds to the plateau modulus (indicated by lines in \cref{Liss} b and d), $G_0=180$ Pa, which is typically determined at much higher frequencies ($De>>1$), indicating that the elastic modulus of the transient micellar network is \textit{constant} across a wider range of lower frequencies than previously thought [Fig. S2(a)]. 
We contrast these observations with the storage modulus, which is frequently taken as a measure of elastic modulus, which has values of $G'(De=0.0625)$= 0.62 Pa and $G'(De=0.25)$= 10.3 Pa [Fig. S7].

The new viscous Lissajous curves from WLMs under LAOS, $\sigma-\dot{\gamma}_{un}$, are shown in \cref{Liss} (f and h). Similar to the $\sigma-\gamma_{rec}$ curves, straight lines with constant slopes are obtained in the small and intermediate amplitudes. This slope is independent of frequency and equal to the zero-shear viscosity, $\eta_0$, determined from steady shearing (straight lines in \cref{Liss}).
The viscosities determined during LAOS and steady shearing are favorably compared in \cref{Liss} (i and j). Over the course of an oscillation at $De=0.0625$, the two protocols yield similar data. Even at the higher frequency ($De=0.25$), the viscosity determined from the LVE and thinning portions in LAOS is consistent with the flow viscosity on the upward sweep. 

The determination of a constant elastic modulus and a constant viscosity in the linear regime across a broad range of frequencies requires reconciliation with the frequency dependence of the dynamic moduli, $G'$ and $G''$. The storage and loss moduli are known to represent the average amount of energy stored and dissipated per unit volume over a cycle of oscillation \cite{Tschoegl:1989kn}, while the elastic modulus presented here is the amount by which the stress increases with an increase in the recoverable strain, or the modulus in a force-extension perspective. 
Having determined the elastic modulus and the viscosity, and the amount of deformation that is recoverable and unrecoverable at each instant, the \textit{instantaneous} energy storage and dissipation rate can be quantified at any time: $W_{s}(t)=\frac{1}{2}G(t)\gamma_{rec}^2(t)$ and $\dot{W}_d(t)=\eta(t)\dot{\gamma}_{un}^2(t)$. Averaging the instantaneous energy storage and dissipation over an oscillation results in metrics that are related to the dynamic moduli: $W_{s,avg}=(\gamma_0^2/4)G'$ and $\dot{W}_{d,avg}=(\gamma_0^2/2)\omega G''$. 
We show that following an energetic analysis, one can transform from the elastic modulus and viscosity to the energetic parameters, $G'$ and $G''$ (Figs. S7 and S8).
The linear-regime dynamic moduli can therefore be obtained \textit{within} a LAOS response if one focuses exclusively on the response to small recoverable strains (Fig. S7). We suggest on this basis that LAOS tests, rather than being treated as being fundamentally different from other tests \cite{Pipkin:1986kf,Ewoldt:2008hdb,Ewoldt:2017ib}, sequentially present information regarding the linear-regime relaxation spectrum and the steady-state flow curve, and how the material transitions between the two states. 

\begin{figure}[h]
	\centering
	\includegraphics[width=0.45\textwidth]{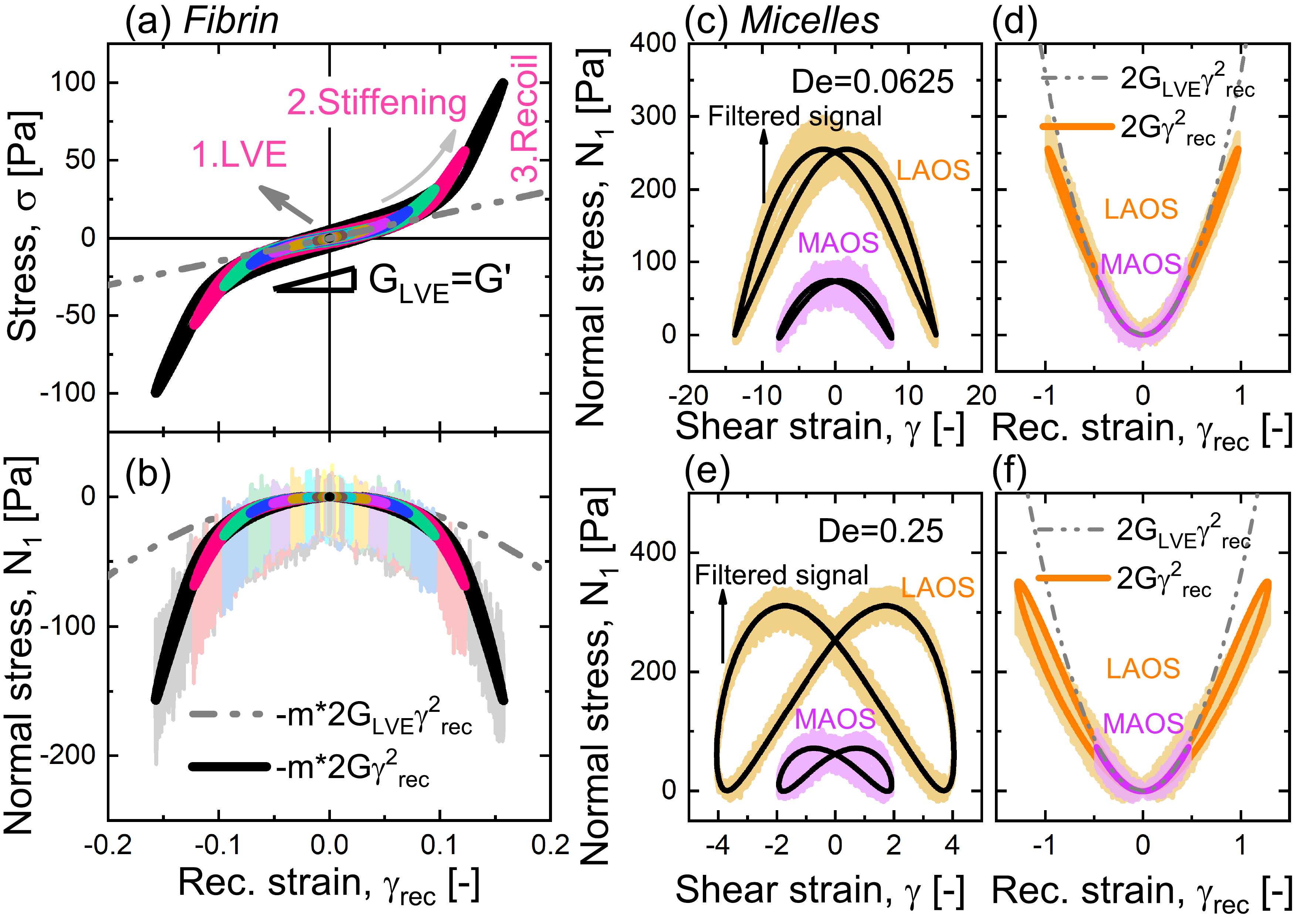}
	\caption{Shear (a) and normal stress (b) are plotted against the recoverable strain from the fibrin.
	Normal stress from WLMs with respect to the strain $\gamma$ and the recoverable strain $\gamma_{rec}$ at $De=0.0625$ (c, d) and $De=0.25$ (e, f). 
	}
	\label{viscosity_N1}
\end{figure}

At instants when the total strain is large (stars in \cref{Liss}), the recoverable strain and the unrecoverable strain rate are minimal, and the WLMs exhibit linear viscoelasticity characterized by $G_0$ and $\eta_0$. Measuring the evolution of the recoverable and unrecoverable strains establishes a clear sequence of processes during a period of oscillation: LVE behavior followed by softening/thinning and recoil, taking place twice per oscillation. 

Unlike the significant differences between the total and recoverable strains in WLMs, the fibrin network shows nearly complete strain recovery (Fig. S3). 
The oscillatory shearing response of fibrin at an imposed frequency of 1 rad/s is presented in \cref{viscosity_N1} (a). Across all deformation amplitudes, when $\mid\gamma_{rec}\mid<\sim$0.05, we observe straight lines with the same slope, indicating a single modulus that is approximately equal to $G'$. This equivalence is because $\gamma_{rec}\approx\gamma$, which is not true for elastic liquids, such as the WLMs.

With increasing recoverable strain, the fibrin stiffens \cite{Munster2013}. Immediately after the recoverable strain reversal, the fibrin recoils to its LVE configuration. Our sequencing of the fibrin responses under oscillatory shearing is in resonance with other studies that account for the unique mechanical properties as a sequence of structural hierarchy with respect to the (recoverable) strain \cite{Onck2005,Piechocka2010,Kang_2009}.

While many studies have focused exclusively on the shear stress response to LAOS, normal stress differences have received little attention \cite{Nam:2010gd,Janmey:2007cp,Seth:2011bj}. A full exposition of the extra stress tensor necessarily involves both normal and shear components. 
Janmey et al. \cite{Janmey:2007cp} studied fibrin networks and saw negative normal stresses under shear. A micromechanical model has been proposed and used to study normal stress differences of soft glasses \cite{Seth:2011bj}. 
Despite these works, a clear link between normal stress differences and other components of the extra stress tensor under LAOS remains incompletely understood. 
In their study of silicone polymers, Benbow and Howells concluded that, ``\textit{the observation of recoverable elastic strain may be taken as a necessary and sufficient condition for observable normal stress} \cite{BENBOW:1961uh}.'' Their conclusion, however, was drawn from steady, uni-directional shearing, and has not been examined under transient shear, nor linked to any structural measure. 

\begin{figure}[h]
	\centering
	\includegraphics[width=0.45\textwidth]{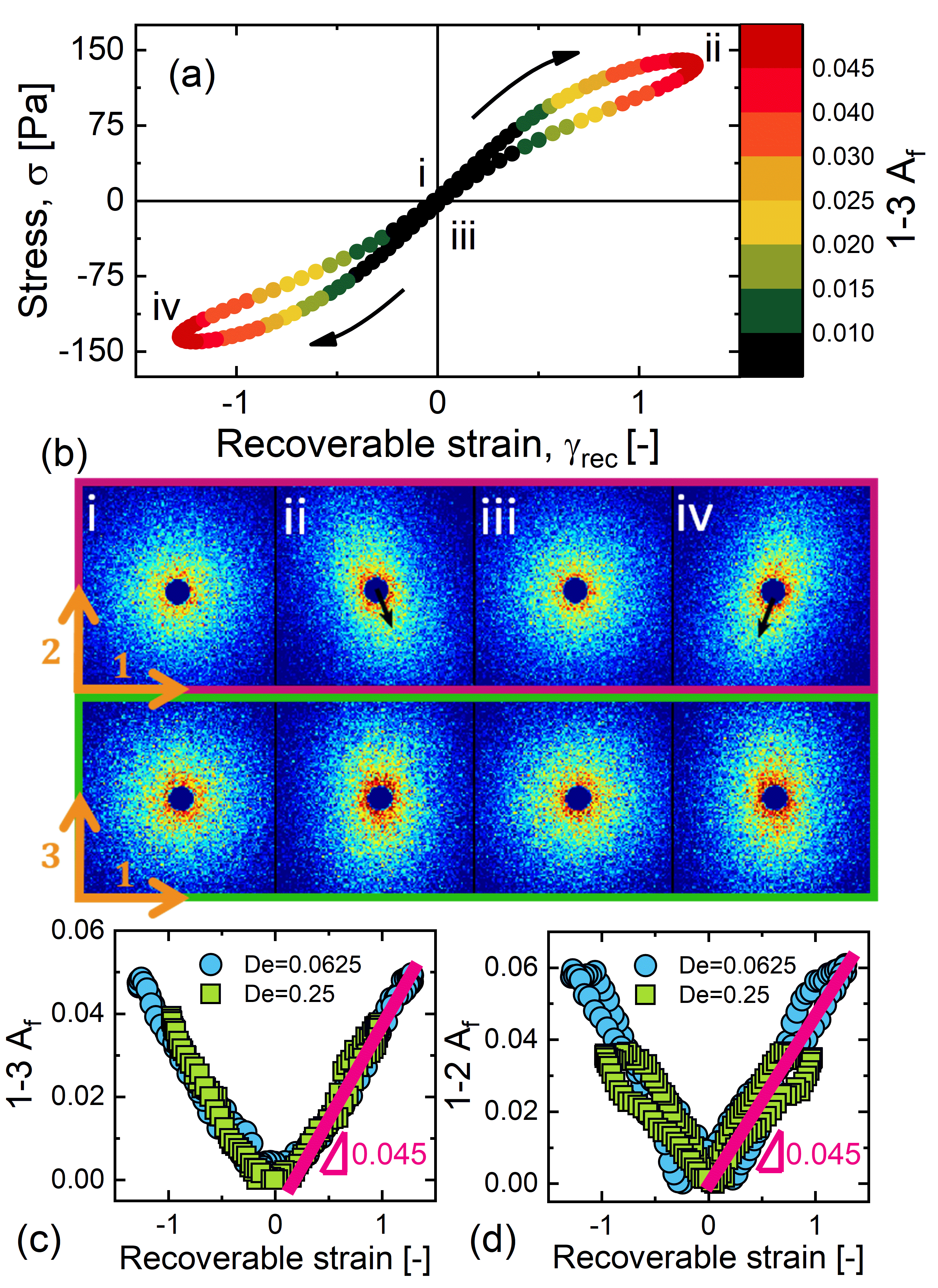}
	\caption{Correlation between macroscopic rheology and microscopic structure. (a) Proposed elastic Lissajous figure ($\sigma-\gamma_{rec}$) coupled with 1-3 alignment factor $A_\text{f}$ denoted by the color scale. (b) 2-D SANS patterns at zero (i and iii) and maximum (ii and iv) recoverable strain in both velocity-gradient and velocity-vorticity planes. Alignment factors in the 1-3 (c) and 1-2 (d) planes are plotted as functions of the recoverable strain, showing the same proportionality constant of 0.045.
	}
	\label{SANS}
\end{figure}

The first normal stress difference, $N_1$, from WLMs is shown in \cref{viscosity_N1} (c) and (e), where the raw response was filtered to remove noise (black lines). When plotted against the total strain, typical bow-tie or butterfly shapes are observed. 
The normal stress trajectory also depends strongly on the imposed amplitude, as the weakly nonlinear response obtained from medium-amplitude (MAOS) and the fully nonlinear response obtained from large-amplitude cases clearly differ. 
When the normal stress differences are plotted in the recoverable strain domain, $N_1-\gamma_{rec}$, there is collapse onto the same parabolic master curve, showing that the MAOS and LAOS cases undergo similar sequences within a cycle of oscillation. 
We therefore observe a quadratic dependence of the normal stress difference on the recoverable strain.

To form a more quantitative understanding of the normal stress difference, we apply an expression derived by Lodge, from his transient network theory of polymers \cite{Lodge:1958iu}. Lodge showed that the ratio of the first normal stress difference to the shear stress under constrained recovery is $N_1/\sigma=2\gamma_{rec}$. We exploit this relationship to provide a description of the first normal stress difference with respect to the recoverable strain, $N_1(t)=2G(t)\gamma_{rec}^2(t)$, where $G(t)$ is the recoverable strain-dependent elastic modulus determined from the $\sigma - \gamma_{rec}$ plots. The calculated $N_1$ values are shown in \cref{viscosity_N1} (d and f) in comparison with the experimental WLM results at the two frequencies. In both cases there is exceptional agreement. Additionally, we show the theoretical prediction from the linear regime, $N_1(t)=2G_{\text{LVE}}\gamma_{rec}^2(t)$ as dashed lines, and note the departure from LVE behavior at $\gamma_{rec}\approx$0.5, in agreement with the shear stress rheology.

The fibrin normal stress is shown in \cref{viscosity_N1}(b) as a function of the recoverable strain. The negative normal stress indicates that the fibrin contracts under shearing \cite{Janmey:2007cp}. Despite having fundamentally distinct structure and properties from WLMs, the magnitude of the normal stress still increases quadratically with the recoverable strain. While it is known that biopolymers do not follow classical rubbery elasticity \cite{Mackintosh1995}, we empirically observe that the normal stress is well described by $N_1(t)=-2mG(t)\gamma_{rec}^2(t)$, where $m$ is found to be 5. Having $m>$1 agrees with the finding that these biopolymers tend to show larger normal stresses than synthetic polymers \cite{Janmey:2007cp,Kang_2009}.
With the results from the two distinct systems, we conclude that the recoverable strain dictates not only the shear stress, but also the complex normal stress. 

The experimentally-determined physical processes observed in the macroscopic rheological responses are mirrored in the microstructural evolution of the WLMs as shown in \cref{SANS} for the $De=0.25$ case for a total strain amplitude of $\gamma_0 = 4$. Shown in \cref{SANS} (a) is the $\sigma - \gamma_{rec}$ figure, with a color scale that reflects the degree of alignment observed in the 2-D SANS patterns shown in \cref{SANS} (b). We quantify the alignment of the micellar segments by defining an alignment factor ($A_\text{f}$) as $A_\text{f}=\int_{0}^{2\pi}I_\text{c}(q^*,\phi)cos(2(\phi-\phi_0))d\phi/\int_{0}^{2\pi}I_\text{c}(q^*)d\phi$, where $I_\text{c}(q^*,\phi)$ is the azimuthal intensity over $q^*$ and $\phi$ is the azimuthal angle with the segmental $q$-range $q^*=0.006$ to 0.03 $1/\AA$. $\phi_0$ represents an orientation angle. The alignment factors in the 1-2 and 1-3 planes are presented as functions of the recoverable strain in \cref{SANS} (c) and (d). 

Contrary to the established view that shear-induced alignment is correlated with shear rate at low frequency and strain at high frequency \cite{Rogers_2012softmatter}, we observe that the alignment of micellar segments is linearly proportional to the magnitude of the recoverable strain regardless of the imposed frequency. Further, the same proportionality coefficient is determined in both the 1-2 and 1-3 planes: $A_f = 0.045\mid{\gamma_{\text{rec}}}\mid$. We therefore build a remarkably consistent physical picture: when the recoverable strain is small, linear viscoelastic responses are elicited, even under LAOS, and the scattering patterns are identical to equilibrium conditions with no alignment of the micellar segments. As the magnitude of the recoverable strain increases, so too do the alignment factor and shear and normal stresses. Even when the modulus begins to drop at large $\mid\gamma_{rec}\mid$ (points ii and iv in \cref{SANS} (a)), the alignment is still linearly dependent on $\mid\gamma_{rec}\mid$.

Biopolymers such as fibrin are known to align and stretch with strain \citep{Vader2009}, exhibiting unique force-extension relationships \cite{Storm2005,Janmey:2007cp}. We have shown that almost all strain is acquired recoverably by fibrin networks. Our results from both the shear-thinning linear entangled micelles and the strain-stiffening fibrin networks therefore show that it is the recoverable strain that provides the basis of accurate nonlinear structure-property relations of soft polymeric materials. 


Access to NGB 30m-SANS was provided by the Center for High Resolution Neutron Scattering, a partnership between the National Institute of Standards and Technology and the National Science Foundation under Agreement No. DMR-1508249. This material is based upon work supported by the National Science Foundation under Grant No. 1727605. We gratefully acknowledge helpful discussions with Michelle Calabrese for the deconvolution protocol, and Connie Hong for preparing the fibrin. Certain commercial instruments or materials are identified in this paper to foster understanding. Such identification does not imply recommendation or endorsement by the National Institute of Standards and Technology, nor does it imply that the materials or equipment identified are necessarily the best available for the purpose.


%

\end{document}



\title{Supplemental information for ``Structure-property relationships via recovery rheology in viscoelastic materials''}

\author{Johnny Ching-Wei Lee}
\affiliation{%
Department of Chemical and Biomolecular Engineering, University of Illinois at Urbana-Champaign, Illinois, USA 61801
}%

\author{Katie M. Weigandt}
\affiliation{%
Center for Neutron Research, National Institute of Standards and Technology, Gaithersburg, MD, USA
}%

\author{Elizabeth G. Kelley}
\affiliation{%
Center for Neutron Research, National Institute of Standards and Technology, Gaithersburg, MD, USA
}%

\author{Simon A. Rogers}%
\affiliation{%
Department of Chemical and Biomolecular Engineering, University of Illinois at Urbana-Champaign, Illinois, USA 61801
}%
\maketitle

\section{Materials}
The studied wormlike micelle solution is composed of 8 wt\% cetylpyridinium chloride (CPCl, Spectrum Laboratory) in 0.5M brine/D$_2$O (Sigma-Aldrich) with a molar ratio of sodium salicylate (NaSal, Sigma-Aldrich) to CPCl of 0.5. 

To prepare the fibrin network, reaction mixtures comprised of 6 mg/mL fibrinogen (Sigma-Aldrich) and 0.2 NIHU/mL thrombin (Sigma-Aldrich), both from human plasma, were mixed in a centrifuge tube. The reaction components were thoroughly mixed with a pipette tip. Subsequently, the mixture was promptly transferred from the tube to the rheometer bottom plate that is preheated at 25 $^\circ$C. We performed the experiments after the system has polymerized for at least 3 hours on the rheometer. To prevent dehydration, the sample is sealed with low-viscosity mineral oil throughout the measurements.

\section{Traditional rheological characterization}

\begin{figure}[h]
	\centering
	\includegraphics[width=0.8\textwidth]{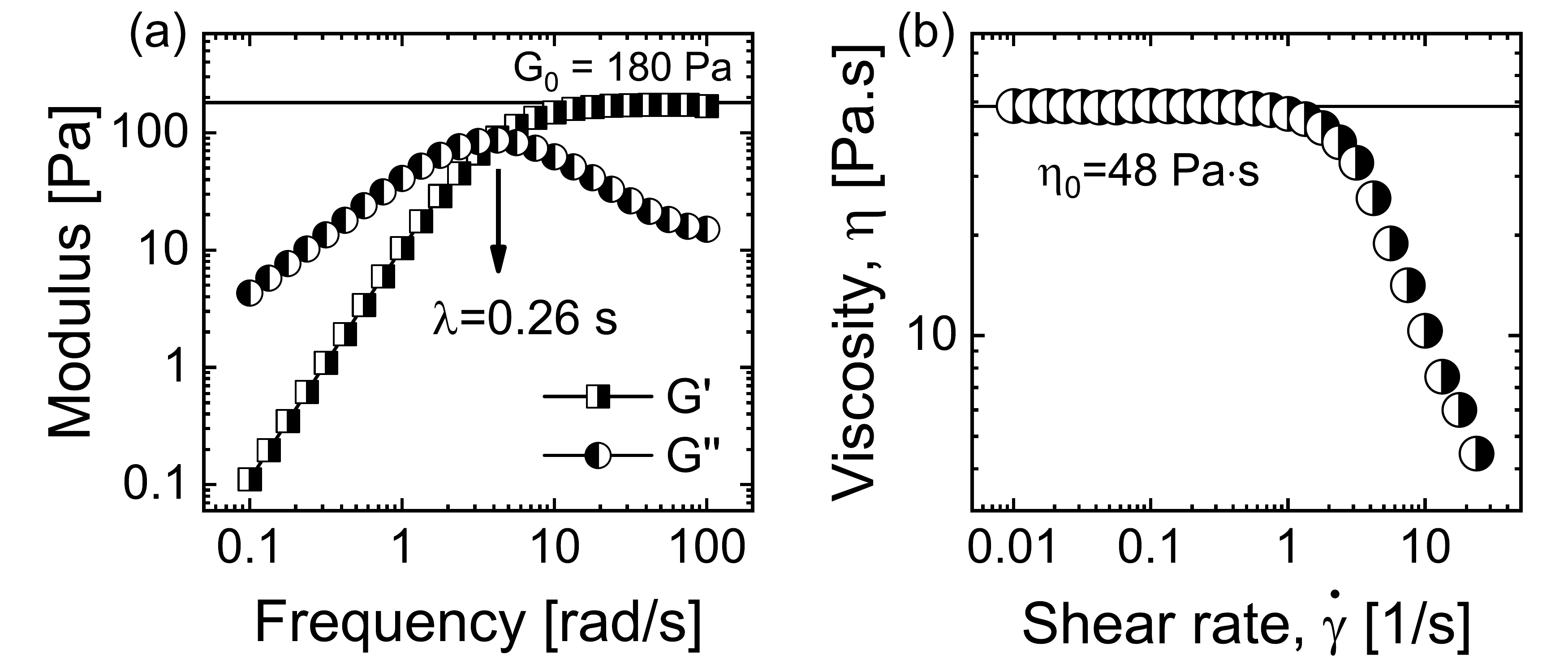}
	\caption{Traditional rheological characterization of the micelle solution. (a) Linear-regime frequency sweep of the dynamic moduli, where the plateau modulus $G_0=180$ Pa and relaxation time $\lambda$=0.26 s are determined. (b) Steady-shear flow curve with the zero-shear viscosity $\eta_0$=48 Pa$\cdot$s. 
	}
	\label{SI_1}
\end{figure}

The frequency dependence of dynamic moduli from the micellar solution is shown in \cref{SI_1}(a), where the plateau modulus $G_0$ and the relaxation time $\lambda$ are 180 Pa and 0.26 s. The steady-shear flow curve is presented in \cref{SI_1}(b), where the zero-shear viscosity $\eta_0$ is 48 Pa$\cdot$s.

\begin{figure}[h]
	\centering
	\includegraphics[width=0.45\textwidth]{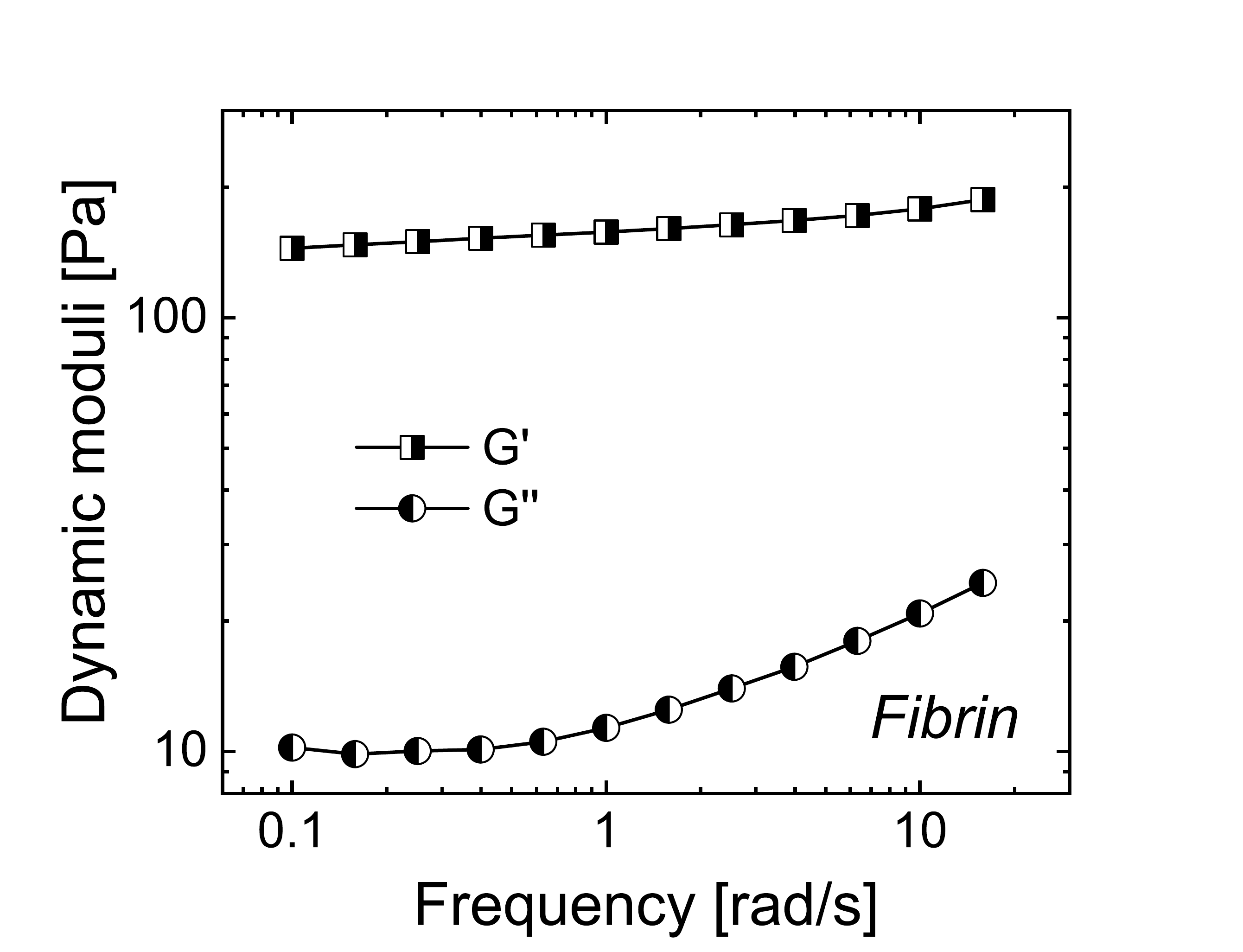}
	\caption{Linear-regime frequency sweep of the dynamic moduli from the fibrin network.  
	}
	\label{SI_fibrin_fre}
\end{figure}

For the fibrin, the linear-regime dynamic moduli are shown as the function of imposed frequency in \cref{SI_fibrin_fre}. Unlike the micelle solution, the fibrin network shows only slight frequency dependence, typical for gel-like network. We therefore show the further experiments at 1 rad/s as an example.

\section{Constrained recovery protocol}

\begin{figure}[h]
	\centering
	\includegraphics[width=0.8\textwidth]{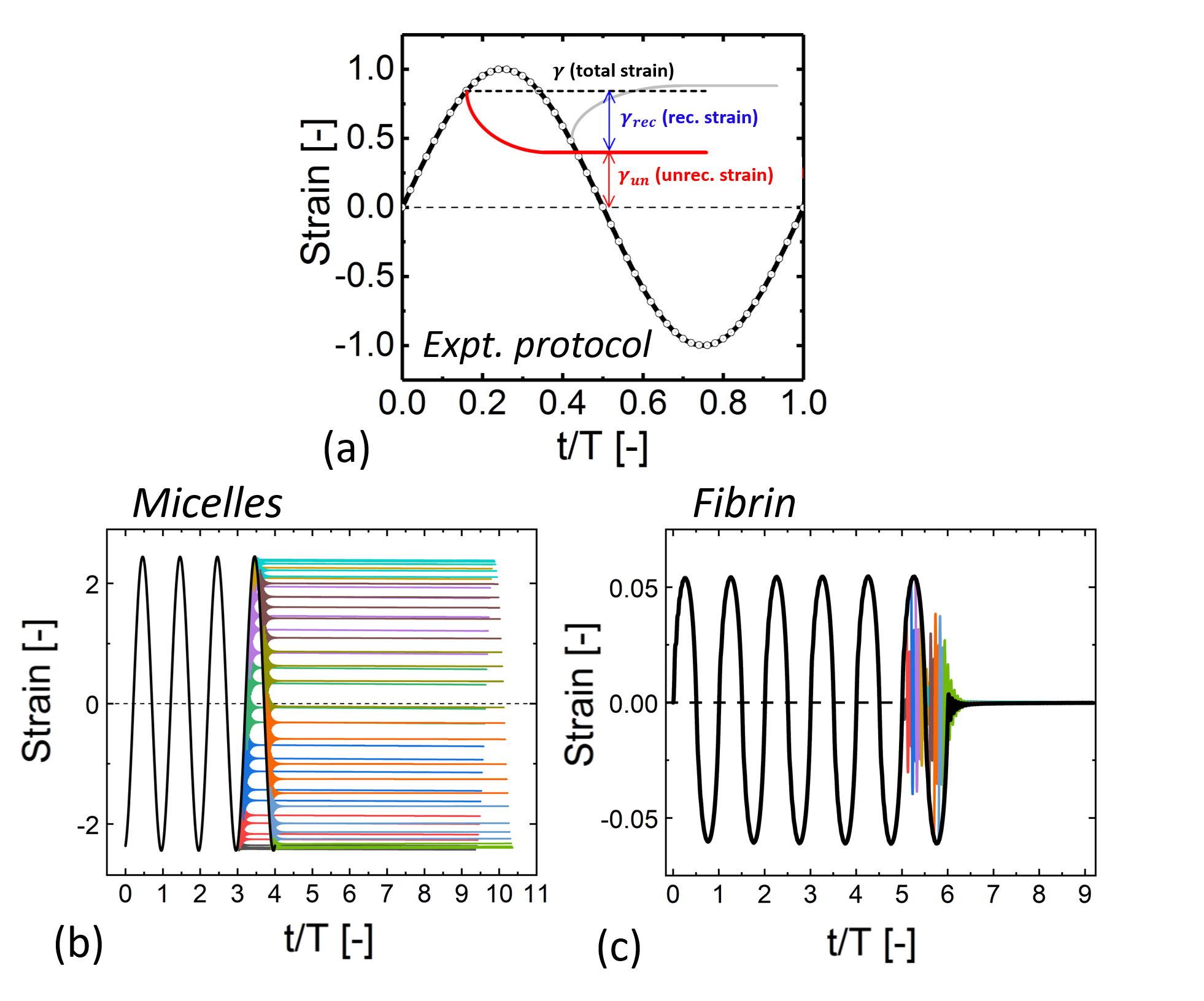}
	\caption{Detailed experimental protocol. (a) A stress-free recovery test is instantaneously imposed at some instant, $t$, after the steady alternating state has been reached. The recoverable and unrecoverable strains are hence determined. Once the recoverable and unrecoverable strains are measured for one instant, the material is allowed to fully relax before another oscillatory period begins and is interrupted by a recovery test at a slightly later relative instant, $t+dt$. An iterative constrained recovery procedure is employed, and the constrained recovery is recorded, at 200 distinct evenly-spaced instants along an oscillation. The raw data following constrained recovery protocol is demonstrated from the micelles (b) and fibrin (c) at 1 rad/s. 
	}
	\label{SI_4}
\end{figure}

A stress-free recovery test is instantaneously imposed at some instant, $t$, after the steady alternating state has been reached. The unrecovered part of the strain, $\gamma_{un}$, is the strain the system ultimately recovers to, and the recoverable strain, $\gamma_{rec}$, is the part of the total strain that is elastically recovered, as shown in \cref{SI_4}(a).
Once the recoverable and unrecoverable strains are measured for one instant, the material is allowed to fully relax before another oscillatory period begins and is interrupted by a recovery test at a slightly later relative instant, $t+dt$. An iterative constrained recovery procedure is employed, and the constrained recovery is recorded, at 200 distinct evenly-spaced instants along an oscillation.

Exemplary constrained recovery results from the micellar and fibrin systems are presented in \cref{SI_4}(b) and (c), respectively. In both the systems, transient ringing phenomena are observed in the initial recovery parts due to the inertio-elastic coupling \cite{Baravian1998}. Shortly after the ringing decays, the systems ultimately recovers to the corresponding unrecoverable strain. 
The recoverable and unrecoverable strains of the micelles vary over the course of an oscillation, while the strain from the fibrin is nearly all recoverable.

\section{Static SANS confirms rod-like regime}

\begin{figure}[h]
	\centering
	\includegraphics[width=0.5\textwidth]{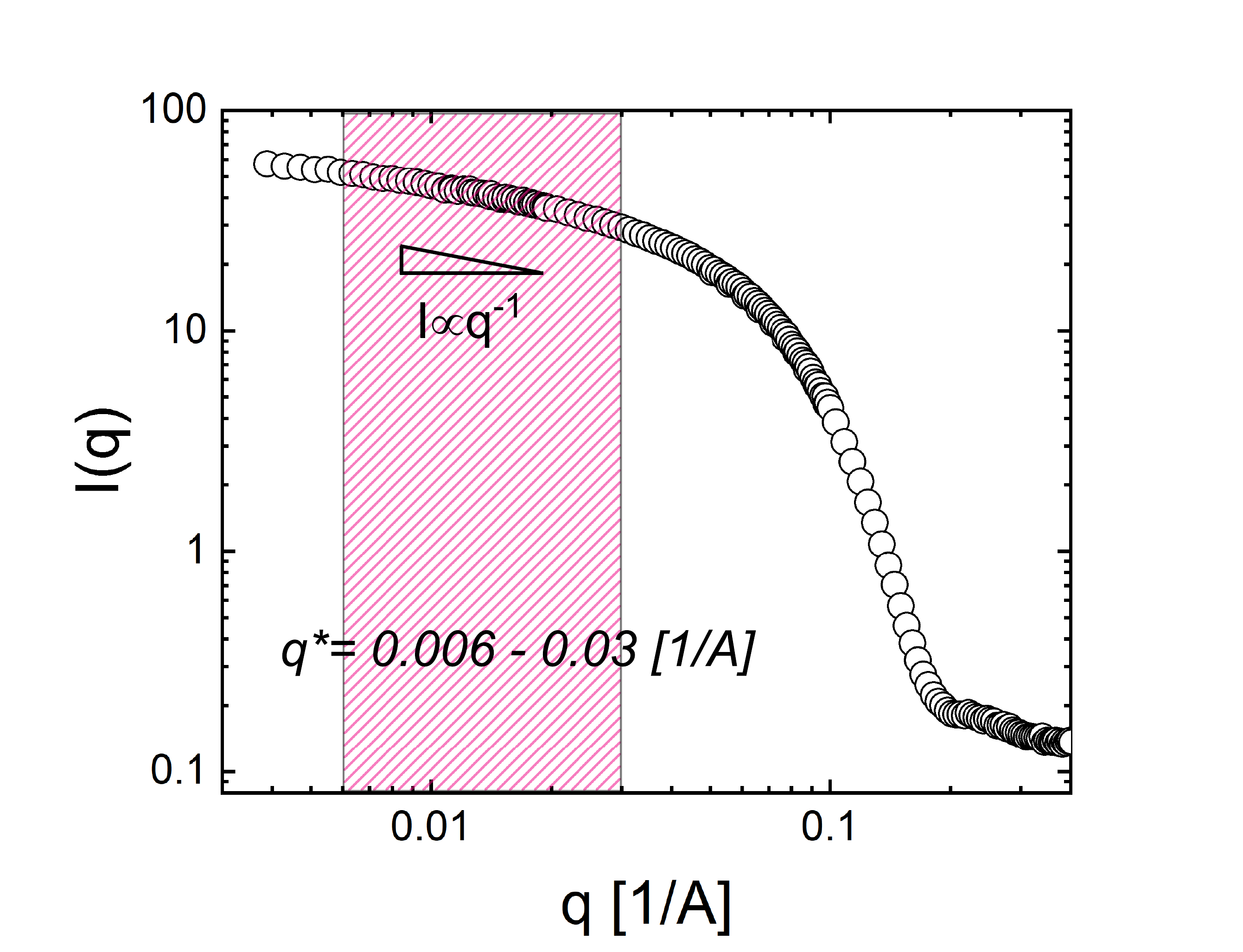}
	\caption{Circular-averaged $I(q)$ versus $q$-vector from the micelle solution. The shaded regime ($q^*$ =0.006 1/$\AA$ to 0.03 1/$\AA$) represents the rod-like scattering regime where $I\propto\ q^{-1}$.
	}
	\label{SI_Iq}
\end{figure}

Circular-averaged SANS intensity $I(q)$ from the micelle solution is displayed in \cref{SI_Iq} as a function of $q$-vector. The shaded region indicates the $q$ range that corresponds to the rod-like scattering $I\propto\ q^{-1}$, $q^*$ =0.006 1/$\AA$ to 0.03 1/$\AA$. This is the $q$-range where the alignment of Kuhn segments are determined.

\section{Elastic modulus $G$ versus the storage modulus $G'$}

\begin{figure}[h]
	\centering
	\includegraphics[width=0.5\textwidth]{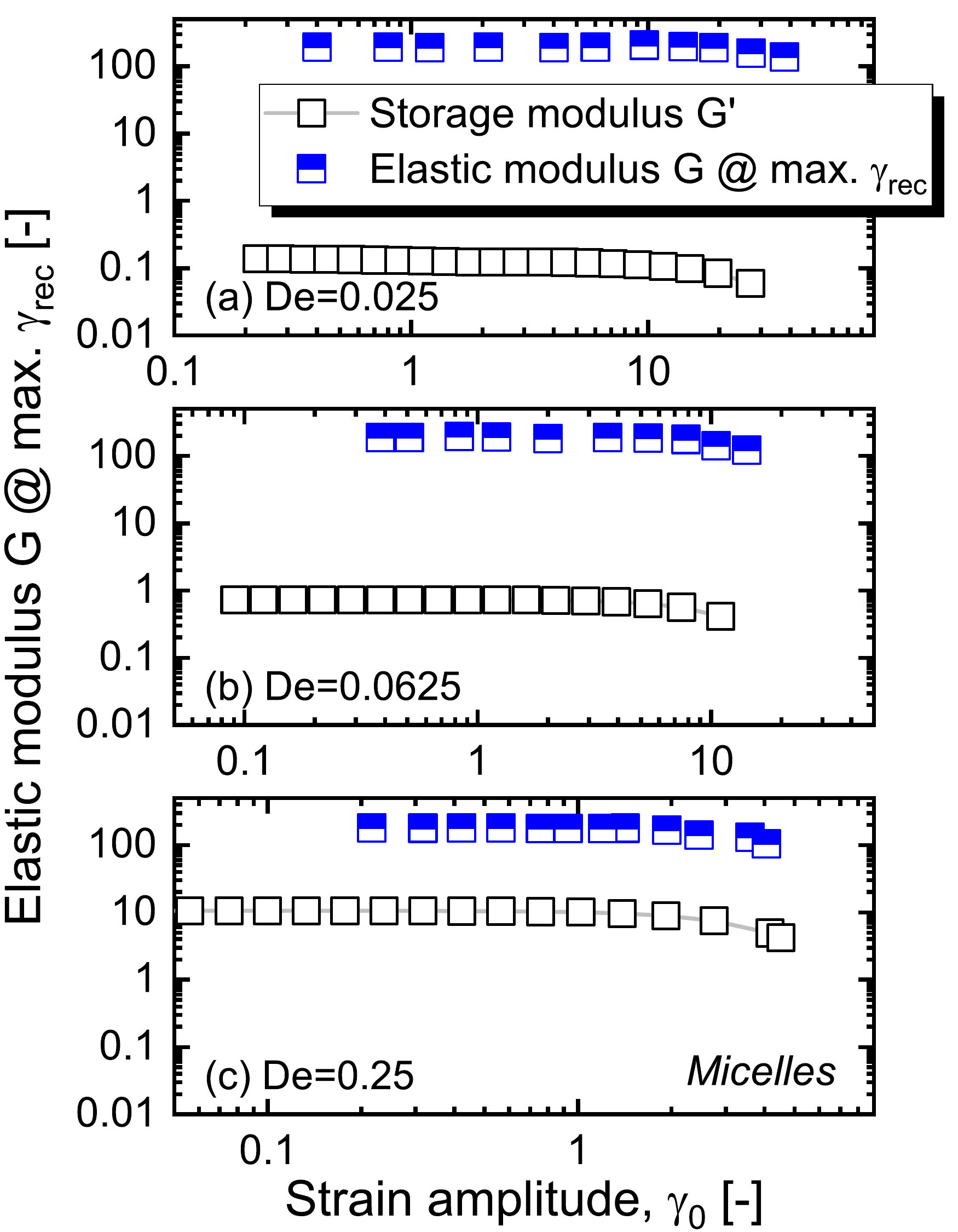}
	\caption{A comparison between the elastic modulus determined at the recoverable strain extreme  $G|_{\gamma_{rec}=max.}$ and the storage modulus $G'$ as a function of imposed amplitude at $De$= 0.025 (a), 0.0625 (b) and 0.25 (c) from the micellar solution.
	}
	\label{SI_elastic_modulus_wlm}
\end{figure}

The storage modulus $G'$ is commonly considered and used as an elastic modulus. However, by definition, it is an \textit{energetic} parameter that represents the average amount of energy stored per unit volume over a cycle of oscillation with units of modulus.
In this section, we contrast the elastic modulus at the recoverable strain extreme, determined as $G|_{\gamma_{rec}=max.}=\sigma/\gamma_{rec}$, to the storage modulus at the corresponding amplitude, from both the micelle and fibrin systems. The elastic modulus $G|_{\gamma_{rec}=max.}$ and the storage modulus $G'$ at $De$= 0.025, 0.0625 and 0.25 from WLMs are shown in \cref{SI_elastic_modulus_wlm} (a), (b) and (c), respectively. In the small amplitudes, while $G|_{\gamma_{rec}=max.}$ and $G'$ both show constant values, they deviate significantly from each other. For instance, at the case of $De$=0.025, the two parameters differ by around three orders-of-magnitude. With increasing imposed frequency, the two parameters get closer due to the increased strain recovery. 
Notably, across a decade of imposed frequency, the critical strain at which the responses become nonlinear as determined by both $G|_{\gamma_{rec}=max.}$ and $G'$ is roughly the same. 

\begin{figure}[h]
	\centering
	\includegraphics[width=0.5\textwidth]{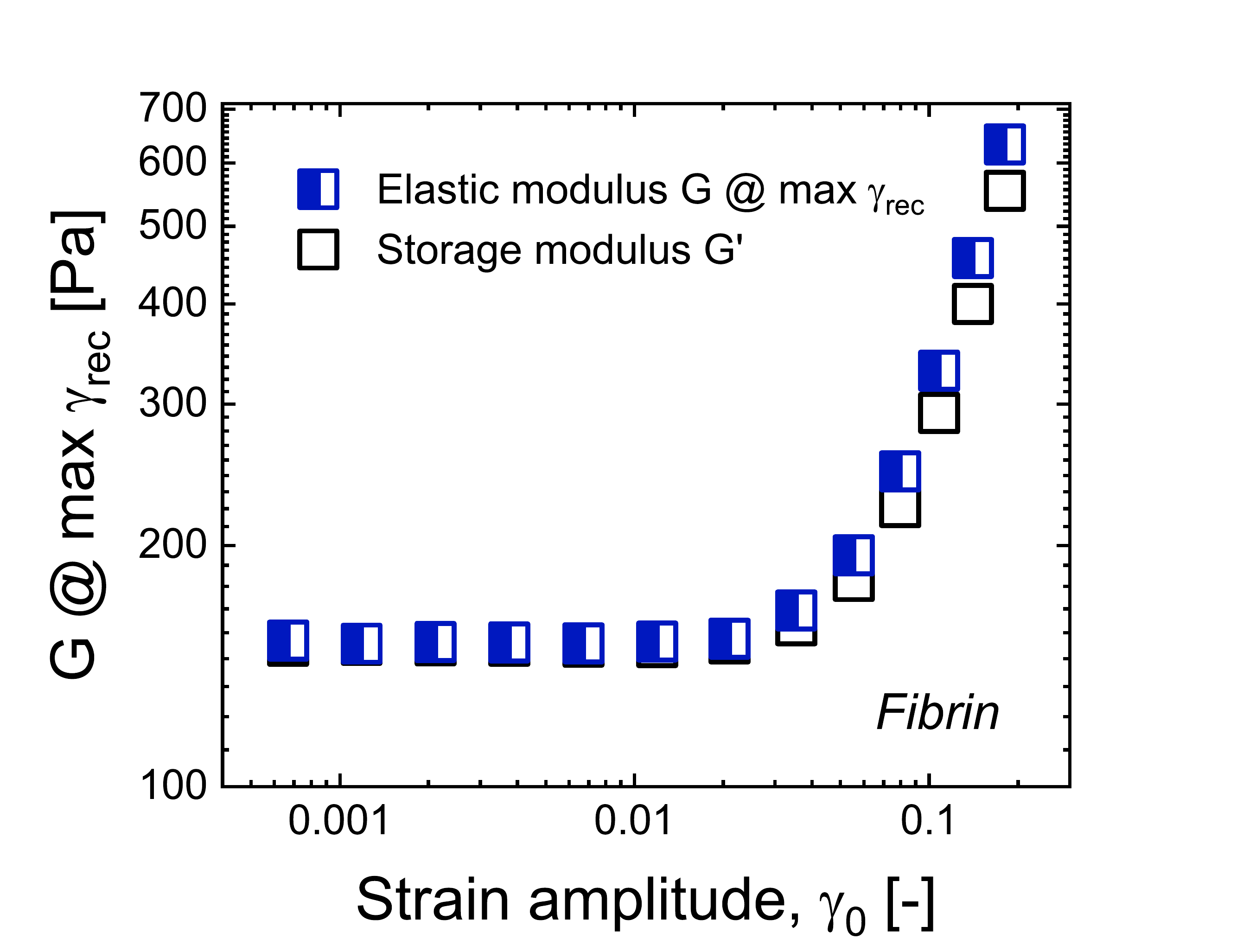}
	\caption{A comparison between the elastic modulus determined at the recoverable strain extreme  $G|_{\gamma_{rec}=max.}$ and the storage modulus $G'$ as a function of imposed amplitude from the fibrin system at 1 rad/s.
	}
	\label{SI_elastic_modulus_fibrin}
\end{figure}

The elastic modulus $G|_{\gamma_{rec}=max.}$ and the storage modulus $G'$ from the fibrin system are presented in \cref{SI_elastic_modulus_fibrin}. The elastic modulus and the storage modulus are identical to each other at small amplitudes. This phenomenon, that the energetic term corresponds to the elastic modulus, is due to the unique properties of fibrin network, that the strain is nearly all recoverable. With the increased amplitudes, the two parameters increase indicating the strain-stiffening features. The storage modulus also gradually deviates from the elastic modulus at the recoverable strain extreme. The storage modulus, despite being an energetic parameter, in this case is a good measure of the average stiffness over a cycle of deformation.

\section{Elastic modulus and viscosity vs. the storage and loss moduli}

\begin{figure}[h]
	\centering
	\includegraphics[width=0.55\textwidth]{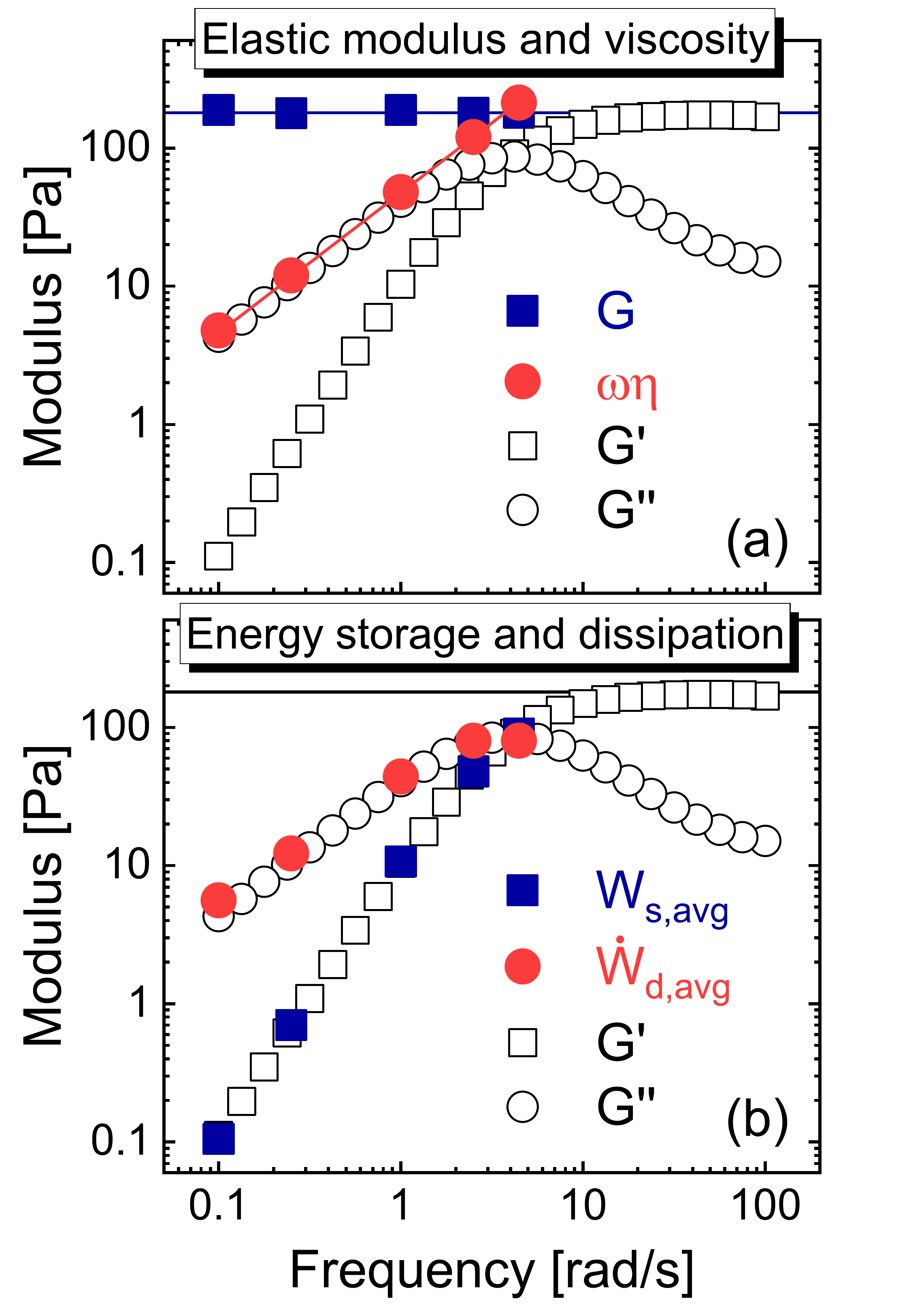}
	\caption{Comparison to the linear-regime dynamic moduli across the frequency $G'(\omega)$ and $G''(\omega)$. (a) The elastic modulus $G$ and viscosity $\eta$ are determined to be constant across the range of frequency, where $G=G_0$=180 Pa and $\eta=\eta_0=48$ Pa$\cdot$s. Note that the constant modulus and viscosity are observed at the small and intermediate amplitudes, as well as the LVE portion of LAOS responses. The determined viscosity is multiplied by frequency $\omega\eta$ for comparison. (b) The average energy storage $W_{s,avg}$ and dissipation rate $\dot{W}_{d,avg}$ over a cycle of oscillation determined from \cref{SIeq1,SIeq2,SIeq3,SIeq4}.
	}
	\label{SI_2}
\end{figure}

We have shown that, for the micellar system, the elastic modulus $G$ and viscosity $\eta$ determined from the recovery rheology deviate significantly from the storage $G'$ and loss $G''$ moduli, as shown in \cref{SI_2}(a). The determination of a constant elastic modulus and viscosity in the linear regime across a broad range of frequencies requires reconciliation with the frequency dependence of the dynamic moduli, $G'(\omega)$ and $G''(\omega)$. The storage and loss moduli are known to represent the average amount of energy stored and dissipated per unit volume over a cycle of oscillation \cite{Tschoegl:1989kn}, while the elastic modulus presented here is the amount by which the stress increases with an increase in the recoverable strain, or the modulus in a force-extension perspective. 
Having determined the elastic modulus and the viscosity, and the amount of deformation that is recoverable and unrecoverable at each instant, the \textit{instantaneous} energy storage and dissipation rate can be quantified at any time: 
\begin{equation}
    W_{s}(t)=\frac{1}{2}G(t)\gamma_{rec}^2(t)
    \label{SIeq1}
\end{equation}
 and 
 \begin{equation}
    \dot{W}_d(t)=\eta(t)\dot{\gamma}_{un}^2(t).
    \label{SIeq2}
 \end{equation}
 Averaging the instantaneous energy storage and dissipation over an oscillation results in metrics that are related to the dynamic moduli: 
 \begin{equation}
    W_{s,avg}=(\gamma_0^2/4)G'
    \label{SIeq3}
 \end{equation}
 and 
 \begin{equation}
    \dot{W}_{d,avg}=(\gamma_0^2/2)\omega G''.
    \label{SIeq4}
 \end{equation}
We show that following the energetic analysis, one can transform from the elastic modulus and viscosity to the energetic parameters, $G'$ and $G''$.
The linear-regime dynamic moduli can therefore be obtained \textit{within} a LAOS response if one focuses exclusively on the small recoverable strain response, as presented in \cref{SI_2}(b). 

\begin{figure}[h]
	\centering
	\includegraphics[width=0.7\textwidth]{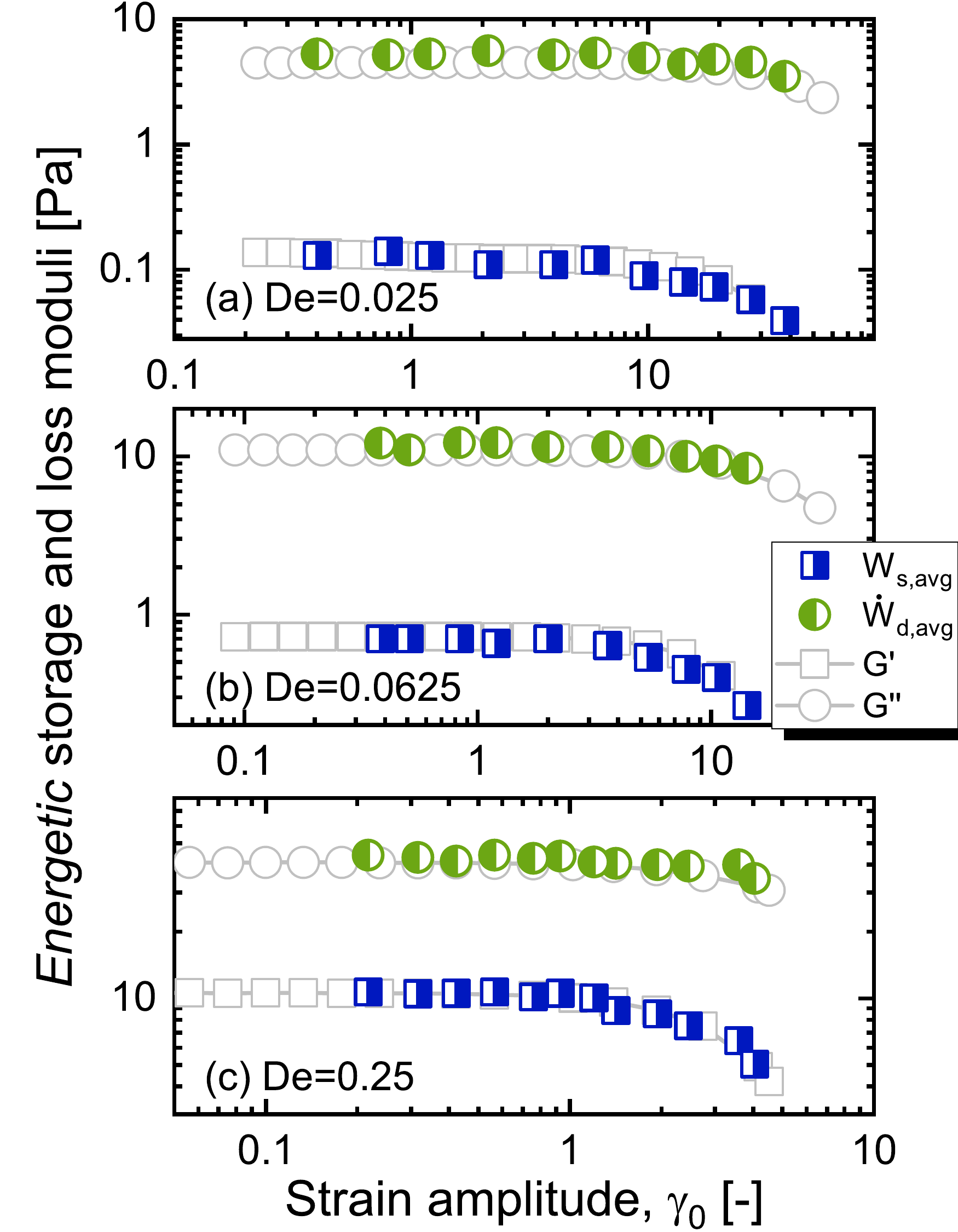}
	\caption{The average energy storage $W_{s,avg}$ and dissipation $\dot{W}_{d,avg}$ are compared with the dynamic moduli across the amplitude $G'(\gamma_0)$ and $G''(\gamma_0)$. 
	}
	\label{SI_3}
\end{figure}

Beyond the linear regime, following the energetic analysis in \cref{SIeq1,SIeq2,SIeq3,SIeq4}, we demonstrate in \cref{SI_3} that the dynamic moduli, $G'(\gamma_0)$ and $G''(\gamma_0)$, can also be accurately determined, with the time-dependent elastic modulus, viscosity, recoverable and unrecoverable strains in the nonlinear regime.

\section{Additional data at a low frequency}

\begin{figure}[h]
	\centering
	\includegraphics[width=0.8\textwidth]{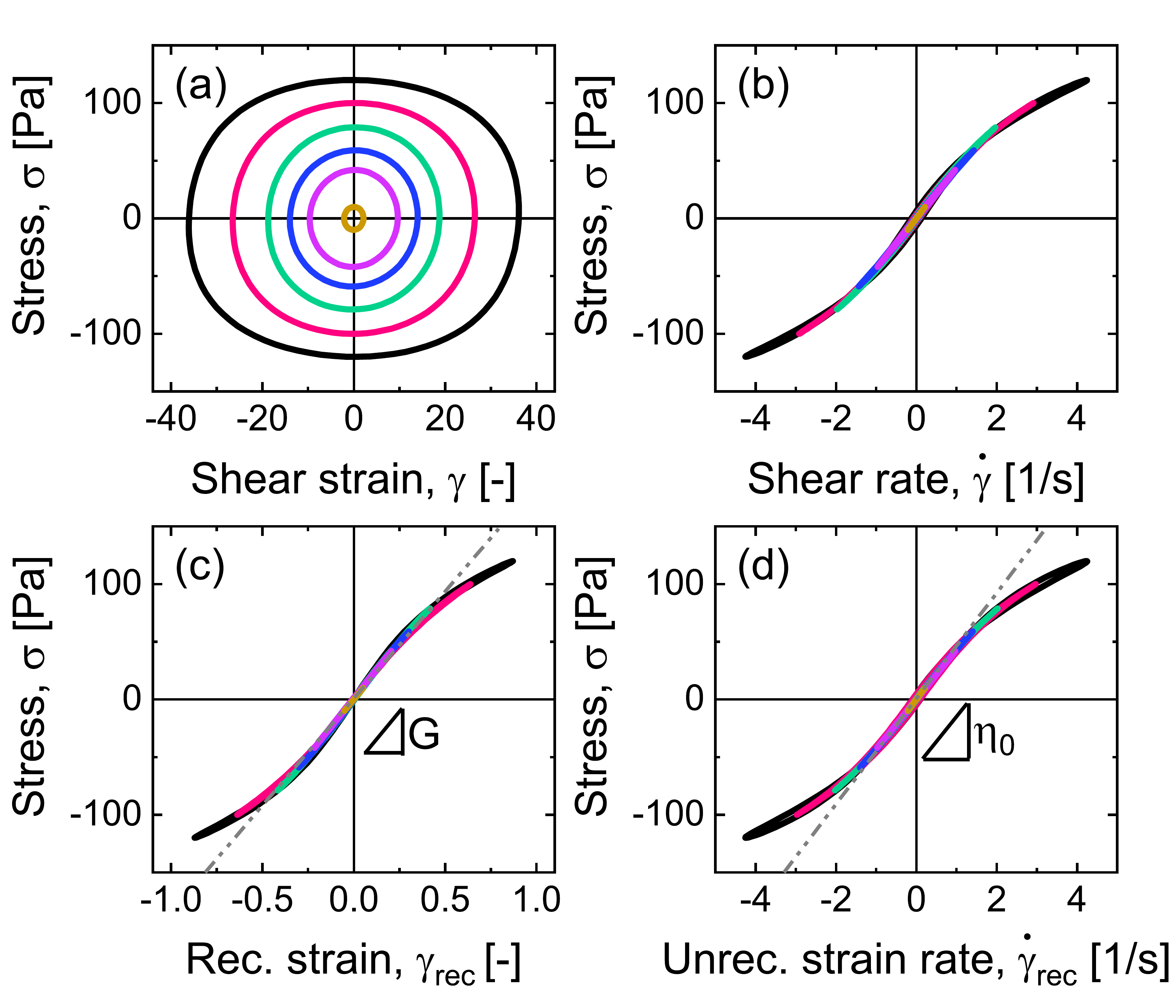}
	\caption{Lissajous curves in the traditional ($\sigma-\gamma$, $\sigma-\dot{\gamma}$) and proposed ($\sigma-\gamma_{rec}$, $\sigma-\dot{\gamma}_{un}$) frames at $De=0.025$ from the micelles. The lines in the proposed elastic and viscous views have the slopes of the plateau modulus $G_0=180 Pa$ and the zero-shear viscosity $\eta_0=48$ Pa$\cdot$s. 
	}
	\label{SI_0p1rads}
\end{figure}

Supplementarily, the Lissajous curves in the traditional ($\sigma-\gamma$, $\sigma-\dot{\gamma}$) and proposed ($\sigma-\gamma_{rec}$, $\sigma-\dot{\gamma}_{un}$) frames at a even lower frequency, $\omega=$0.1 rad/s ($De$=0.025), are shown in \cref{SI_0p1rads}. We again observe that the plateau modulus $G_0$ and zero-shear viscosity $\eta_0$ can be clearly obtained at the small recoverable strain, in the presentations in \cref{SI_0p1rads}(c) and (d). The same sequence, LVE followed by softening/thinning and recoil, is manifested along a cycle of oscillation, across the investigated frequencies.

%